\documentclass[aps,pra,twocolumn,superscriptaddress,a4paper]{revtex4}
\usepackage{amsmath}
\usepackage{graphicx}

\begin{document}

\title{Simulation of high-spin Heisenberg models in coupled cavities}

\author{Jaeyoon Cho}
\affiliation{Department of Physics and Astronomy, University College London, Gower St., London WC1E 6BT, UK}

\author{Dimitris G. Angelakis}
\affiliation{Centre for Quantum Technologies, National University of Singapore, 2 Science Drive 3, Singapore 117542}
\affiliation{Science Department, Technical University of Crete,
Chania, Crete, Greece, 73100}

\author{Sougato Bose}
\affiliation{Department of Physics and Astronomy, University College London, Gower St., London WC1E 6BT, UK}

\date{\today}

\begin{abstract}
We propose a scheme to realize the Heisenberg model of any spin in an arbitrary array of coupled cavities. Our scheme is based on a fixed number of atoms confined in each cavity and collectively applied constant laser fields, and is in a regime where both atomic and cavity excitations are suppressed. It is shown that as well as optically controlling the effective spin Hamiltonian, it is also possible to engineer the magnitude of the spin. Our scheme would open up an unprecedented way to simulate otherwise intractable high-spin problems in many-body physics.
\end{abstract}

\pacs{}

\maketitle

\newcommand{\abs}[1]{\left|#1\right|}
\newcommand{\ket}[1]{\left|#1\right>}
\newcommand{\bra}[1]{\left<#1\right|}

The Heisenberg spin model has played a crucial role as a basic model accounting for the magnetic and thermodynamic natures of many-body systems. Despite extensive investigations, however, many aspects of the model are still largely unexplored both analytically and numerically, especially for the cases of higher spins. The main difficulty in the numerical treatment originates from the fact that the Hilbert-space dimension blows up exponentially as the number of spins increases. As Feynman first noted \cite{feynman82}, this difficulty would be overcome in terms of quantum simulation based on precisely controlled quantum systems. Realization of quantum simulation, expected in a near future, will mark a milestone towards the realization of sophisticated quantum computation.

In the context of quantum information processing, a qubit is identical to an $s=\frac12$ spin, and in a few implementations, such as the arrays of Josephson junctions \cite{bruder93} or quantum dots \cite{loss98}, the spin-chain Hamiltonian naturally emerges from the spin-like coupling between qubits, albeit with limited control of the coupling constants. On the other hand, in optical lattices, perturbative evolution with respect to the Mott-insulator state can be described by an effective spin-chain Hamiltonian \cite{duan03,garcia-ripoll04}. This approach has its own merit in that the spin-coupling constants can be optically controlled to a great extent. An alternative approach, recently under active investigation, is to use the array of coupled cavities, which are ideally suited to addressing individual spins \cite{hartmann06,greentree06,angelakis07,hartmann07,rossini07,paternostro07,cho07a,li08,kay08}. In this approach, a spin is represented by either polaritons or hyperfine ground levels. The former, proposed in Refs.~\cite{angelakis07} and \cite{kay08}, allows a stronger spin-spin coupling than the latter, but lacks the optical control of the coupling. On the other hand, the latter, proposed in Ref.~\cite{hartmann07}, retains the optical controllability, but relies on rapid switching of optical pulses and the consequent Trotter expansion, which unavoidably involves additional errors and makes error-free implementation more difficult. More importantly, the question of simulating chains of higher spins, which may have a completely different phase diagram, remains open. In some sense, these are more important to simulate because unlike spin-$\frac12$ chains, they do not have exact analytical solutions for a wide range of parameters including the XXX case, except for special kinds of models \cite{affleck89}. Additionally, going to higher spins should make perturbative spin-wave theory more accurate, whose predictions can be tested.

In this paper, we propose a scheme to realize the anisotropic (XXZ) or isotropic (XXX) Heisenberg spin model of any spin in an arbitrary array of coupled cavities. Our scheme is experimentally feasible in that simply applying a small number of constant laser fields suffices for our purpose. If the number of lasers is increased, the individual constants of the spin Hamiltonian are controlled more flexibly. Most of all, a strong advantage of our scheme is that the magnitude of the spin itself can be engineered arbitrarily. This advantage contrasts with all the earlier schemes mentioned above including those for optical lattices, in which the spin is fixed in nature mostly as $s=\frac12$ ($s=1$ in Ref.~\cite{garcia-ripoll04}). $s>\frac12$ spin chains exhibit fascinating physics that $s=\frac12$ spin chains can not have. A well-known example is Haldane's conjecture that antiferromagnetic Heisenberg integral-spin chains have a unique disordered ground state with a finite excitation gap, whereas half-integral-spin chains are gapless \cite{haldane83a,haldane83b}. Our scheme could be used to prepare a ground state, for example, through an adiabatic evolution, and measure its excitation gap and spin correlation functions \cite{garcia-ripoll04}. {The advantage of our scheme is, however, more apparent when we consider higher-spin problems intractable with any previous method. One of the intriguing examples is the quantum spin dynamics of the ferric wheels such as $\text{Fe}_6$ and $\text{Fe}_{10}$ composed of $s=\frac52$ spins \cite{meier01}. These systems would be simulable with a relatively small number (6 or 10) of cavities.} Spin chains also play an important role as a quantum channel for short-distance quantum communication \cite{bose07}. The property of an $s=1$ antiferromagnetic spin chain as a quantum channel strongly depends on its phase \cite{romero-isart07}. In some phases, it provides an efficient channel, outperforming that of a ferromagnetic chain. It has been also shown that a ground state with an excitation gap, as is the case for the spin-1 chain, can serve as a more efficient quantum channel \cite{hartmann06b}. {The ground state of the antiferromagnetic spin-1 chain also establishes the so-called localizable entanglement between two ends, which can be extracted by measuring every intermediate spins \cite{verstraete04} and then used for quantum communication. It is hard to demonstrate these schemes in optical lattices owing to the difficulty of addressing individual sites \cite{cho07b}.} Although the idea of communicating using spin chains is ultimately meant for solid-state applications, our system can serve as a preliminary test for comparing and contrasting the performance of various spin-$s$ chains.

\begin{figure}
\centerline{\includegraphics[height=.16\textheight]{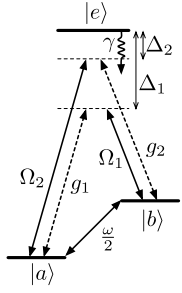}}
\caption{Involved atomic levels and transitions. Both transitions $\ket{a}\leftrightarrow\ket{e}$ and $\ket{b}\leftrightarrow\ket{e}$ are coupled to the same cavity mode with coupling rates $g_1$ and $g_2$ and with detunings $\Delta_1$ and $\Delta_2$, respectively. Two laser fields with Rabi frequency $\Omega_{1}$ and $\Omega_{2}$ are also applied with detunings $\Delta_1$ and $\Delta_2$, respectively, and the transition between ground levels $\ket{a}$ and $\ket{b}$ is driven with Rabi frequency $\omega/2$ by Raman lasers. $\gamma$ denotes the atomic spontaneous decay rate.}
\label{fig:simple}
\end{figure}

We use two ground levels of a three-level atom to represent an $s=\frac12$ spin (in a rotated basis, as will be seen later). We start by recalling that in terms of two states $\ket\downarrow$ and $\ket\uparrow$ of one atom, the $s=\frac12$ spin is described in terms of operators $s^Z=\frac12(\ket\uparrow\bra\uparrow-\ket\downarrow\bra\downarrow)$, $s^+=\ket\uparrow\bra\downarrow$, and $s^-=\ket\downarrow\bra\uparrow$. Our starting point is an observation that if there are $M$ identical atoms, one can straightforwardly define total spin operator $S^Z=\sum_{j=1}^Ms_j^Z$ with $S^\pm=\sum_{j=1}^Ms_j^\pm$ ($j$ is the index for the atoms), by which the atoms represent $S=\frac M2$, $\frac M2-1$, and so on. Keeping this in mind, let us consider a coupled array of identical cavities {in an arbitrary geometry}, each of which contains $M$ identical single atoms. {We employ the Dicke-type model, in which every atom in a cavity interacts with the cavity mode with the same coupling strength \cite{dicke54}.} Let us first consider a simple case, as depicted in Fig.~\ref{fig:simple}. Let us denote by $\ket\psi_{jk}$ the state $\ket\psi$ of the $k$th atom in the $j$th cavity. In the rotating frame, the Hamiltonian reads
\begin{equation}
\begin{split}
H=&\sum_{j}\left[e^{i\Delta_1t}\Omega_1\Lambda_j^{eb}+e^{i\Delta_2t}\Omega_2\Lambda_j^{ea}+h.c.\right]\\
+&\sum_{j}\left[(e^{i\Delta_1t}g_1\Lambda_j^{ea}+e^{i\Delta_2t}g_2\Lambda_j^{eb})a_j+h.c.\right]\\
+&\sum_{j}\frac\omega2(\Lambda_j^{ab}+\Lambda_j^{ba})-\sum_{\left<j,k\right>}J(a_j^\dagger a_{k}+a_ja_{k}^\dagger),
\end{split}
\end{equation}
where $\Lambda_j^{xy}=\sum_{k=1}^{M}(\ket{x}\bra{y})_{jk}$ ($x,y=a,b,e$), $a_j$ is the annihilation operator for the $j$th cavity mode, $\Delta_j$ is the corresponding detuning, $\Omega_j$ and $\frac\omega2$ are the corresponding Rabi frequencies of the classical fields, $g_j$ is the corresponding atom-cavity coupling rate, and $J$ is the inter-cavity hopping rate of photons. Both the transitions are coupled to the same cavity mode. The transition between $\ket{a}$ and $\ket{b}$ is induced by two-photon Raman transition using far-detuned lasers. {Here, $\left<j,k\right>$ represent nearest neighbor pairs.} For now, we ignore the spontaneous decay rate $\gamma$ of the atom.

Before we proceed, it is instructive to write down our parameter regime:
\begin{gather}
\frac{g_1^2}{\Delta_1}=\frac{g_2^2}{\Delta_2},\label{eq:cond1}\\
{\Delta_j},{\Delta_1-\Delta_2}\gg \sqrt{\frac{M}{2}}g_j\gg J\gtrsim\abs{\Omega_j},\label{eq:cond2}\\
{M\frac{g_1^2}{\Delta_1}}\sim\abs{M\frac{g_1^2}{\Delta_1}\pm\omega}\sim\abs{\omega}\gg2J\label{eq:cond3}.
\end{gather}
The condition~(\ref{eq:cond1}) can be fulfilled with conventionally used alkali-metal atoms, such as rubidium and caesium. For example, one may choose ground hyperfine levels $\ket{F=1,m_F=-1}$ and $\ket{F=2,m_F=-1}$ of a ${}^{87}\text{Rb}$ atom to represent $\ket{a}$ and $\ket{b}$, respectively, and use $\sigma^+$-polarized light, for which $g_1>g_2$. The detuning $\Delta_j$ is then comparable to the hyperfine splitting between the two levels. Although there are multiple excited levels, their contributions can be summed up and denoted by single parameters in what follows. The other conditions~(\ref{eq:cond2}) and (\ref{eq:cond3}) can be satisfied simultaneously when $\sqrt{\frac{M}{2}}g_j/{\Delta_j}\gg J/\sqrt{\frac{M}{2}}g_j$. For example, our scheme works well in case ${\Delta_j}/1000\sim\sqrt{\frac{M}{2}}g_j/100\sim J$, which is allowed by strong atom-cavity coupling.

Our regime is chosen so that the excitation of the atom or the cavity photon is suppressed (condition~(\ref{eq:cond2})), while the communication between atoms is mediated by virtual cavity photons. The first step is to adiabatically eliminate the excited state using the conventional method, by which the effective Hamiltonian is given by
\begin{equation}
\begin{split}
H=&-\sum_{j}\frac{g_1^2}{\Delta_1}\left(\Lambda_j^{aa}+\Lambda_j^{bb}\right)a_j^\dagger a_j\\
&-\sum_{j}\left[\left(\mu_1\Lambda_j^{ba}+\mu_2\Lambda_j^{ab}\right)a_j+h.c.\right]\\
&+\sum_{j}\frac\omega2(\Lambda_j^{ab}+\Lambda_j^{ba})-\sum_{\left<j,k\right>}J(a_j^\dagger a_{k}+a_ja_{k}^\dagger),
\end{split}
\end{equation}
where $\mu_j=\frac{g_j\Omega_j^*}{\Delta_j}$. Now let us introduce spin operators in a rotated basis
\begin{equation}
\left\{\ket\uparrow=\frac{1}{\sqrt2}(\ket a+\ket b)\text,\ket\downarrow=\frac{1}{\sqrt2}(\ket a-\ket b)\right\}
\end{equation}
to represent an $s=\frac12$ spin. Note that these are the eigenstates of $\frac\omega2(\ket{a}\bra{b}+\ket{b}\bra{a})$. The underlying idea is to apply the Raman lasers with Rabi frequency $\frac\omega2$ constantly, introducing a fixed amount of energy splitting $\abs\omega$ between the two spin states. The total spin is then defined in terms of the operators
\begin{align}
S_{j}^{Z}=\sum_{k=1}^{M}s_{jk}^{Z}\text{ and }S_{j}^{\pm}=\sum_{k=1}^{M}s_{jk}^{\pm},
\end{align}
where $s_{jk}^{Z}=\frac12(\ket{\uparrow}\bra{\uparrow}-\ket{\downarrow}\bra{\downarrow})_{jk}$, $S_{jk}^{+}=(\ket\uparrow\bra\downarrow)_{jk}$, and $s_{jk}^{-}=(\ket\downarrow\bra\uparrow)_{jk}$. The total spin is given by $S_{j}^{2}=(S_{j}^{Z})^{2}+\frac12(S_{j}^{+}S_{j}^{-}+S_{j}^{-}S_{J}^{+})$. If $M$ is even (odd), the atoms represent integral (half-integral) spins up to $\frac{M}{2}$. The atomic operators are now written as $\Lambda_j^{\downarrow\downarrow}=\sum_{k}s_{jk}^-s_{jk}^+=\frac{M}{2}-S_j^Z$, $\Lambda_j^{\uparrow\uparrow}=\sum_{k}s_{jk}^+s_{jk}^-=\frac{M}{2}+S_j^Z$, $\Lambda_j^{\uparrow\downarrow}=S_j^+$, and $\Lambda_j^{\downarrow\uparrow}=S_j^-$. Substituting these operators, the Hamiltonian in the rotating frame reads
\begin{equation}
\begin{split}
H=&-\sum_{j}\left[\left\{e^{i\lambda t}\mu_{12}^+S_j^{Z}+e^{i(\lambda+\omega)t}\frac{\mu_{12}^-}{2}S_j^{+}\right.\right.\\
&~~~~~~~~~~\left.\left.-e^{i(\lambda-\omega)t}\frac{\mu_{12}^-}{2}S_j^-\right\}a_j+h.c.\right]\\
&-\sum_{\left<j,k\right>} J(a_j^\dagger a_{k}+a_j a_{k}^\dagger),
\end{split}
\label{eq:inter}
\end{equation}
where $\lambda=M\frac{g_1^2}{\Delta_1}$ and $\mu_{12}^\pm=\mu_1\pm\mu_2$. Note that in view of conditions~(\ref{eq:cond2}) and (\ref{eq:cond3}), the effective Rabi frequency $\abs{\sqrt{\frac{M}{2}}\mu_{12}^\pm}$ is much smaller than $\lambda$, $\abs{\lambda\pm\omega}$, and $\abs\omega$. This allows us to make use of the adiabatic elimination once more. We extend the method in Ref.~\cite{james07} to keep up to the third order terms and take only the subspace with no cavity photon. Simple algebra as in Ref.~\cite{james07} yields the final Heisenberg spin Hamiltonian
\begin{equation}
\begin{split}
H=&\sum_{j}\left[A(S_j)^2+B(S_j^Z)^2+CS_j^Z\right]\\
&-\sum_{\left<j,k\right>}\left[D(S_j^XS_{k}^X+S_j^YS_{k}^Y)+ES_j^ZS_{k}^Z\right],
\end{split}
\label{eq:hamil}
\end{equation}
where $A=\frac{\lambda}{\lambda^2-\omega^2}\frac{\abs{\mu_{12}^-}^2}{2}$, $B=\frac{\abs{\mu_{12}^+}^2}{\lambda}-\frac{\lambda}{\lambda^2-\omega^2}\frac{\abs{\mu_{12}^-}^2}{2}$, $C=-\frac{\omega}{\lambda^2-\omega^2}\frac{\abs{\mu_{12}^-}^2}{2}$, $D=\frac{J}{2}(|{\frac{\mu_{12}^-}{\lambda+\omega}}|^2+|{\frac{\mu_{12}^-}{\lambda-\omega}}|^2)$, $E=2J|{\frac{\mu_{12}^+}{\lambda}}|^2$. This Hamiltonian already covers a wide range of parametric regimes for the Heisenberg spin model, although individual control of the parameters is limited owing to their mutual dependency. Interestingly, the Hamiltonian also contains the single-ion anisotropy $(S_j^Z)^2$, which is of essential importance in high-spin cases \cite{botet83}, whereas it is merely a meaningless constant in the spin-$\frac12$ case.

\begin{figure}
\centerline{\includegraphics[height=.16\textheight]{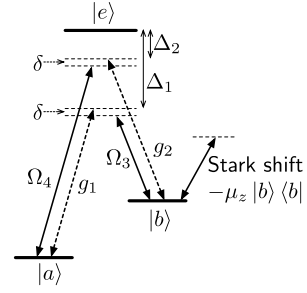}}
\caption{Additional lasers to get full control of the spin Hamiltonian. These lasers are applied in addition to the set up of Fig.~\ref{fig:simple}.}
\label{fig:full}
\end{figure}

Full control of the individual parameters is allowed by bringing in more lasers, shown in FIG.~\ref{fig:full}, in addition to the set up of FIG.~\ref{fig:simple}. The classical fields with Rabi frequency $\Omega_3$ and $\Omega_4$, which are applied with an additional detuning $\delta\sim\lambda$, make a similar contribution to the effective Hamiltonian as those with $\Omega_1$ and $\Omega_2$. This can be reflected in Hamiltonian~(\ref{eq:inter}) by adding the same terms with $\lambda$ and $\mu_{12}^\pm$ replaced by $\lambda-\delta$ and $\mu_{34}^\pm$, respectively, where $\mu_{34}^\pm=\mu_3\pm\mu_4$, $\mu_3=\frac{g_1\Omega_3^*}{2}(\frac{1}{\Delta_1}+\frac{1}{\Delta_1+\delta})$, and $\mu_4=\frac{g_2\Omega_4^*}{2}(\frac{1}{\Delta_2}+\frac{1}{\Delta_2+\delta})$. The Stark shift $-\mu_z\ket{b}\bra{b}$ induced by another far-detuned laser field (using a different level and polarization) results in adding $\sum_{j}\frac{\mu_z}{2}(e^{i\omega t}S_j^++e^{-i\omega t}S_j^-)$ in Hamiltonian~(\ref{eq:inter}). Recall that in our previous derivation, we have adjusted $\{0,\lambda,\lambda\pm\omega\}$ so that they are distinct in frequency with the similar frequency spacing (condition~(\ref{eq:cond3})), thereby causing each summation in the Hamiltonian~(\ref{eq:inter}) to contribute independently to the constants in the final Hamiltonian~(\ref{eq:hamil}). We adjust $\omega$, $\lambda$, $\lambda\pm\omega$, $\lambda-\delta$, and $\lambda-\delta\pm\omega$ in the same spirit. For the ease of presentation, let us take a particular situation where
$\omega>0$, $\lambda=3\omega$, and $\lambda-\delta=-6\omega$,
although this is not a necessary condition. We then obtain the same Hamiltonian~(\ref{eq:hamil}) with parameters given by
$A=\frac{1}{\lambda}(\frac{9}{16}{\abs{\mu_{12}^-}^2}-\frac{9}{35}{\abs{\mu_{34}^-}^2})$,
$B=\frac{1}{\lambda}({\abs{\mu_{12}^+}^2}-\frac{9}{16}{\abs{\mu_{12}^-}^2}-\frac{1}{2}{\abs{\mu_{34}^+}^2}+\frac{9}{35}{\abs{\mu_{34}^-}^2})$,
$C=\frac{1}{\lambda}(\frac32{\abs{\mu_z}^2}-\frac{3}{16}{\abs{\mu_{12}^-}^2}-\frac{3}{70}{\abs{\mu_{34}^-}^2})$,
$D=\frac{J}{\lambda^2}(\frac{45}{32}\abs{{\mu_{12}^-}}^2+\frac{333}{1225}\abs{{\mu_{34}^-}}^2)$, and
$E=\frac{J}{\lambda^2}(2\abs{{\mu_{12}^+}}^2+\frac{1}{2}\abs{{\mu_{34}^+}}^2)$.
Note that $C$ is determined independently thanks to the term ${\abs{\mu_z}^{2}}/{\lambda}$, while other terms are also determined freely. Hence, this parameter set covers any anisotropic or isotropic Heisenberg spin models, with the single-ion anisotropy turned on or off. 

The ground state of the spin Hamiltonian could be prepared by the adiabatic method, as described in Ref.~\cite{garcia-ripoll04}. Although the Hamiltonian~\eqref{eq:hamil} looks similar to a ferromagnetic one ($D,E>0$), one can also simulate the antiferromagnetic spin Hamiltonian, since other parameters $A$, $B$, and $C$ can be adjusted to have any sign. This can be easily seen by noting that if the parameters $A$, $B$, and $C$ are adjusted to be $-1$ times those of a desired antiferromagnetic Hamiltonian, the Hamiltonian is equivalent to the antiferromagnetic one up to global factor $-1$, hence with an inverted energy spectrum. Consequently, the adiabatic preparation, starting from an antiparallel spin configuration~\cite{garcia-ripoll04}, ends up with the highest energy state, which in fact is the ground state of the corresponding antiferromagnetic Hamiltonian.

Although atomic excitation is heavily suppressed, the main source of decoherence in our system is the spontaneous decay of atoms. In relation to the effective spin model, the atomic spontaneous decay results in depolarization of the spins. This effect can be accounted for by considering a conditional Hamiltonian $H_C=H-\frac{i}{2}\sum_{j}({\gamma_A'}\Lambda_j^{aa}+{\gamma_B'}\Lambda_j^{bb})$, where the effective decay rates are approximately given by $\gamma_A'=\gamma\left(\frac{\abs{\Omega_1}^2}{\Delta_1^2}+\frac{\abs{\Omega_3}^2}{(\Delta_1+\delta)^2}\right)$ and $\gamma_B'=\gamma\left(\frac{\abs{\Omega_2}^2}{\Delta_2^2}+\frac{\abs{\Omega_4}^2}{(\Delta_2+\delta)^2}\right)$, assuming other lasers are sufficiently detuned and thus make a negligible contribution to the decay. In particular, if $\Omega_j$s are chosen in such a way that the two contributions are balanced, i.e., $\gamma_A'=\gamma_B'=\gamma'$, the depolarization is nearly independent of the spin state. In this case, the conditional Hamiltonian is approximately given by $H_C=H-iNM\frac{\gamma'}{2}$, where $N$ is the number of cavities. Consequently, the state of the system at time $t$ may be written as $\rho(t)=e^{-NM\gamma't}\rho_Q(t)+(1-e^{-NM\gamma't})\rho_M$, where $\rho_Q(t)$ is the desired quantum state evolved by the spin Hamiltonian and $\rho_M$ is the fully mixed state. This property is useful for testing condensed-matter theories, since even under depolarization, the quantum nature retained in the coherent portion $\rho_Q(t)$ could be observed over a time scale $\sim1/NM\gamma'$. One requirement is that the spin-spin coupling rate multiplied by $(M/2)^2$ should be much larger than the global decoherence rate. Reminding that the coupling rate is given by $\sim2J{\abs{\mu_j}^2}/{\lambda^2}\sim({J}/{M})({\Omega_j}/{\sqrt{M/2}g_j})^2$, we require
$\gamma\ll\frac{J}{2N}\bigl(\frac{\Delta_j}{\sqrt{M/2}g_j}\bigr)^2$.
Since $\Delta_j\gg\sqrt{M/2}g_j$ from condition~(\ref{eq:cond2}), this requirement can be met for a moderate $N$ if $J\gtrsim\gamma$ is satisfied along with our previous assumption of strong atom-cavity coupling. Note, however, that testing Haldane's conjecture for higher-spin chains is more demanding, since the lowest excitation gap is expected, from its asymptotic behavior, to decrease rapidly with increasing $M$, while the spin correlation length increases rapidly \cite{haldane83a}. There are various micro-cavity technologies under active development which are expected to fall into our regime of strong atom-cavity coupling \cite{spillane05}, such as superconducting microwave cavities \cite{majer07}, photonic bandgap microcavities \cite{hennessy07}, and microtoroidal cavities \cite{aoki06}. These models would be also suited to having a fixed number of atoms in a cavity, by virtue of the progress in the micro-fabrication techniques. For example, coupling two superconducting qubits with a single cavity mode in a well-controlled way has been demonstrated recently \cite{majer07}, which suggests the viability of the proposed way of engineering spins \cite{note}. An important point is that contrary to the system-specific ideas, our scheme relies on a general model, which would be available in a wide range of current or future systems.

This work was supported by the Korea Research Foundation Grant (KRF-2007-357-C00016) funded by the Korean Government (MOEHRD). SB thanks the Engineering and Physical Sciences Research Council (EPSRC) UK for an Advanced Research Fellowship and for support through the Quantum Information Processing IRC (GR/S82176/01) and the Royal Society and the Wolfson Foundation.

\end{document}